\documentstyle[11pt,aaspp4]{article}

\received{24 November 1997}

\slugcomment{\bf to appear in The Astrophysical Journal}

\lefthead{Pahre et al.}
\righthead{Surface Brightness Fluctuations with HST}

\begin{document}

\def\etal{{\emph{et al.}}}
\def\I814bar{\overline{I}_{F814W}} 
\def\Io814bar{\overline{I}_{F814W,0}} 
\def\MI814bar{\overline{M}_{I_{F814W}}} 
\def\Ibar{\overline{I}_{\rm C}} 
\def\MIbar{\overline{M}_{I_{\rm C}}} 
\def\Im814{I_{F814W}} 
\def\DM{(m-M)_0} 
\def\simlt{\lower.5ex\hbox{$\; \buildrel < \over \sim \;$}}
\def\simgt{\lower.5ex\hbox{$\; \buildrel > \over \sim \;$}}

\title{Detection of Surface Brightness Fluctuations in NGC~4373 Using the
	Hubble Space Telescope$^1$}

\author{
Michael~A.~Pahre\altaffilmark{2,3,4},
Jeremy~R.~Mould\altaffilmark{5},
Alan~Dressler\altaffilmark{6},
Jon~A.~Holtzman\altaffilmark{7},
Alan~M.~Watson\altaffilmark{8},
John~S.~Gallagher~III\altaffilmark{8},
Gilda~E.~Ballester\altaffilmark{9},
Christopher~J.~Burrows\altaffilmark{10},
Stefano~Casertano\altaffilmark{11},
John~T.~Clarke\altaffilmark{9},
David~Crisp\altaffilmark{12},
Richard~E.~Griffiths\altaffilmark{11},
Carl~J.~Grillmair\altaffilmark{13},
J.~Jeff~Hester\altaffilmark{14},
John~G.~Hoessel\altaffilmark{8},
Paul~A.~Scowen\altaffilmark{14},
Karl~R.~Stapelfeldt\altaffilmark{13},
John~T.~Trauger\altaffilmark{13},
and
James~A.~Westphal\altaffilmark{15}
}

\altaffiltext{1}{Based on observations with the NASA/ESA
        Hubble Space Telescope obtained at the Space Telescope Science
        Institute, which is operated by the Association of Universities for
        Research in Astronomy Inc., under NASA contract NAS~5--26555.}

\altaffiltext{2}{Palomar Observatory,  California Institute of Technology, 105-24 Pasadena, CA 91125 }

\altaffiltext{3}{Present address:  Harvard-Smithsonian Center for Astrophysics, 60 Garden Street, Mail
	Stop 20, Cambridge, MA \, 02138; mpahre@cfa.harvard.edu }

\altaffiltext{4}{Hubble Fellow.}

\altaffiltext{5}{Mount Stromlo and Siding Spring Observatories, Institute of Advanced Studies, 
	Australian National University, Private Bag, Weston Creek Post Office, ACT 2611, Australia; jrm@merlin.anu.edu.au}

\altaffiltext{6}{Observatories of the Carnegie Institution of Washington, 813 Santa Barbara Street, 
	Pasadena, CA  91101; dressler@ociw.edu}

\altaffiltext{7}{Lowell Observatory, Mars Hill Road, Flagstaff, AZ  86001}

\altaffiltext{8}{Department of Astronomy, University of Wisconsin -- Madison, 475 N. Charter St., Madison, WI 53706}

\altaffiltext{9}{Department of Atmospheric and Oceanic Sciences, University of Michigan, 2455 Hayward, Ann Arbor, MI 48109}

\altaffiltext{10}{Space Telescope Science Institute, 3700 San Martin Drive, Baltimore, MD 21218}

\altaffiltext{11}{Department of Astronomy, Johns Hopkins University, 3400 N. Charles St., Baltimore, MD 21218}

\altaffiltext{12}{Jet Propulsion Laboratory, 4800 Oak Grove Drive, Mail Stop 241--105, Pasadena, CA 91109}

\altaffiltext{13}{Jet Propulsion Laboratory, 4800 Oak Grove Drive, Mail Stop 183--900, Pasadena, CA 91109}

\altaffiltext{14}{Department of Physics and Astronomy, Arizona State University, Tyler Mall, Tempe, AZ 85287}

\altaffiltext{15}{Division of Geological and Planetary Sciences, California Institute of Technology, Pasadena, CA 91125}


\begin{abstract}

Surface brightness fluctuations (SBF) have been detected for three elliptical
galaxies---NGC~3379 in the Leo group, NGC~4406 in the Virgo cluster, and NGC~4373 
in the Hydra--Centaurus supercluster---using marginally--sampled, deep images taken 
with the Planetary Camera of the WFPC--2 instrument on the Hubble Space Telescope (HST).  
The power spectrum of the fluctuations image is well--fit by an empirical model of the
point--spread function (PSF) constructed using point sources identified in the field.
The previous ground--based SBF measurements for NGC~3379 and NGC~4406 are recovered,
thereby demonstrating the capability of the Planetary Camera of WFPC--2 to measure
distances using the SBF technique despite the marginal sampling of the images.
The residual variance due to unresolved sources in all three galaxies is only 
2--5\% of the detected fluctuations signal, which confirms the advantage of HST 
imaging in minimizing the uncertainty of this SBF correction.
Extensive consistency checks, including an independent SBF analysis using an alternate 
software package, suggest that our internal uncertainties are $< 0.02$~mag.
The fluctuations magnitude for NGC~4373 is $\I814bar = 31.31 \pm 0.05$~mag, 
corresponding to a distance modulus of $\DM = 32.99 \pm 0.11$.
This implies a peculiar velocity for this galaxy of $415 \pm 330$~km~s$^{-1}$, which is
smaller than derived from the $D_n$--$\sigma$ relation.
These results demonstrate the power of the post--repair HST to measure distances to
elliptical galaxies at significant redshifts using the SBF technique.

\end{abstract}



\section{Introduction \label{intro} }

The measurement of surface brightness fluctuations (SBF) in early--type galaxies 
(\cite{tonry88}; \cite{jacoby92}) is an important development in the investigation 
of the extragalactic distance scale.
The rms scatter in galaxy distance measurements is $\sim 0.18$~mag, according to 
a recent, large survey (\cite{tonry97}), demonstrating that this technique 
is an accurate secondary distance estimator.
A variety of investigations have followed in the optical (\cite{tal89}; \cite{tal90}; 
\cite{tonsch90}; \cite{tonry91}; \cite{simard94}; \cite{sodemann95}; \cite{tonry97}),
near-infrared (\cite{lupton93}; \cite{pahre94}; \cite{jensen96}), and
using HST/WFPC--2 (\cite{ajhar97}; \cite{lauer97}).

The velocity field in the Local Supercluster has been mapped for the first time with 
significant signal--to--noise (S/N) by Tonry (1995\markcite{tonry95}) using ground--based SBF.
Similar ground-based optical observations at distances greater than $v=3000$ or
$4000$~km~s$^{-1}$, however, are limited by the ability to identify the globular 
cluster system (see Tonry \etal\ 1990).
This is a result of the residual variance contributed by unidentified globular
clusters swamping the SBF signal.
For this reason, observations at larger distances have utilized the inversion of the 
method, i.e. removing the stellar SBF signal (due to the galaxy light) in order to study 
the globular cluster systems from the remaining SBF signal (e.g. \cite{blakeslee95}; 
\cite{blakeslee97}).
In order to measure distances of galaxies at $v > 4000$~km~s$^{-1}$ using the SBF method
requires either deep exposures on a very large telescope or substantially--improved 
resolution (or both) in order to characterize the globular cluster system accurately
for each galaxy.

Exploring the local velocity field outside the Local Supercluster will
benefit significantly from the improved angular resolution available from
the Hubble Space Telescope (HST). 
To understand the opportunity provided by Wide--Field/Planetary Camera--2 (WFPC--2) 
for this purpose, the Investigation Definition Team (IDT) has carried out some pilot 
observations in NGC~4373, an elliptical galaxy in the Hydra--Centaurus supercluster. 
This galaxy was selected as a good candidate for SBF observations due to its distance, 
which at $cz\sim 3625$~km~s$^{-1}$ puts it at or beyond the limit of ground-based 
optical SBF measurements.
Its location in the Centaurus region on the front-side of the ``Great Attractor'' 
(e.g. in the models of \cite{lb88}) makes this galaxy additionally interesting.

It was unknown if the Planetary Camera of WFPC--2 could measure SBF accurately, as 
the pixel size only marginally--samples the PSF in the relevant filter F814W 
(roughly corresponding to the $I_{\rm C}$--band).
In this paper, we will demonstrate the feasibility of measuring SBF using HST 
by analyzing images of two relatively nearby elliptical galaxies---NGC~3379 in the 
Leo group and NGC~4406 in the Virgo cluster---and comparing the results with
ground--based SBF measurements.
Then the SBF analysis will be applied to NGC~4373 to measure its distance,
thereby showing the power of HST to extend this secondary distance indicator to larger
distances.

\section{Data Reductions \label{reductions} }

Images of galaxy NGC~4373 were obtained through the F814W filter using the
WFPC--2 on the HST as part of a GTO program.
There were five different exposures taken:  $2 \times 1900$~s, $2 \times 1100$~s,
and $500$~s.
Additional images of galaxies NGC~3379 in the Leo group and NGC~4406 in the Virgo
cluster, also taken through the F814W filter, were obtained from the HST archive.
All of the data were obtained using the inverse gain of $7$~e$^{-}$~DN$^{-1}$ electronics.
The NGC~3379 and NGC~4406 exposures utilized the pipeline processing from the HST archive,
while the NGC~4373 were separately processed in a similar way---including ADC 
correction, bias subtraction, dark subtraction (using super and delta dark frames), 
and flattening.
The frames were then shifted by integer pixel values (to prevent correlating the noise 
in adjacent pixels) and combined using various cosmic--ray rejection algorithms.  
There were two separate pointings observed for NGC~4373 that were intentionally offset by an 
integer number of pixels in order that the images could be reregistered without 
resampling.
The chosen algorithm for combining the images was the STSDAS/CRREJ task, 
although there was very little difference between the result of this and the 
IRAF/IMCOMBINE task.
The FWHM of the PSF in the final images ranged from $\sim 1.7$~pixel for NGC~4373
to $\sim 2.0$~pixel for the other two galaxies.

The zero-point used for the subsequent analysis of NGC~4373 was obtained from 
Holtzman \etal\ (1995b\markcite{holtzman95b}) for transforming the F814W onto $\Im814$ using the color
$(V-I_{\rm C}) = 1.23$ (to be consistent with \cite{ajhar97}) produces a calibration of:
\begin{equation}
\Im814 = 21.546 - 2.5 \log {\rm DN} + 2.5 \log ( {t \over {\rm s} } ) ,
\end{equation}
where $t$ is the integration time.
Note that we have not included an additional $0.1$~mag into this calibration (as was done
by \cite{ajhar97}) since we will normalize the PSF flux to an aperture of $r=0.5$~arcsec 
(see \S~\ref{sbf-analysis-consistency} below), although this will not affect the final result.
Despite the second--order color term that enters into this equation, 
the resultant zero--point only varies by $< 0.01$~mag across the
relevant range of observed color $1 < (V-I_{\rm C}) < 1.4$~mag.

The 100$\mu$m cirrus emission in the region towards and around the three galaxies 
was obtained for $3 \times 3$ grids centered on the galaxies with $\sim 8$~arcmin 
spacing using the IRAS m.eps.
Obvious point sources were visually excluded.
The conversion to extinction is then 
\begin{equation}
A_B ({\rm mag}) = {F_{100} \over 14 \pm 2 {\rm~MJy/ster} } 
\end{equation}
(\cite{laureijs94}).
The relative extinction was assumed to be $A_{F814W} = 0.479 A_B$ (\cite{holtzman95b})
and $E_{V-I} = 0.301 A_B$ (\cite{cardelli89}).
The $A_B$ extinction derived for NGC~4373 is $0.11$~mag larger than was assumed by 
Faber \etal\ (1989\markcite{faber89}).

\begin{figure}
\epsscale{0.9}
\plotone{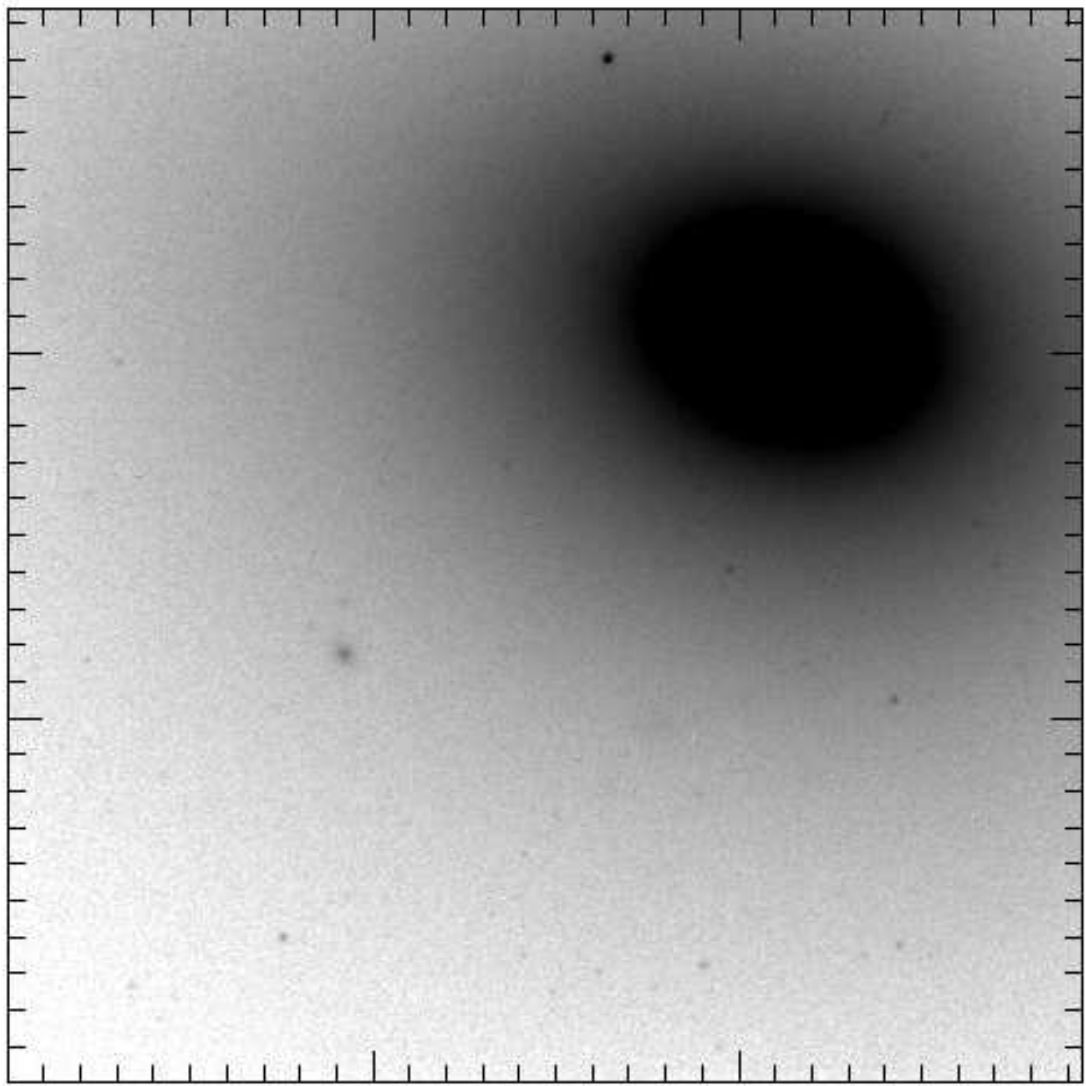}
\caption{
The Planetary Camera grey-scale image of NGC~4373 displayed with a logarithmic stretch.
The small tickmarks on both axes are spaced by 1~arcsec, while the large tickmarks are
spaced by 10~arcsec.
\label{n4373-greyscale}
}
\end{figure}

The exposure times, extinction estimates, and the observed and dereddened colors
for the three galaxies are listed in Table~\ref{table1}.
\placetable{table1}
A greyscale representation of the Planetary Camera image of NGC~4373 is shown 
in Figure~\ref{n4373-greyscale}.
\placefigure{n4373-greyscale}

\section{Surface Brightness Fluctuations Analysis \label{sbf-analysis} }

\subsection{Fitting a Model to the Galaxies \label{sbf-analysis-model} }

\begin{figure}
\epsscale{0.6}
\plotone{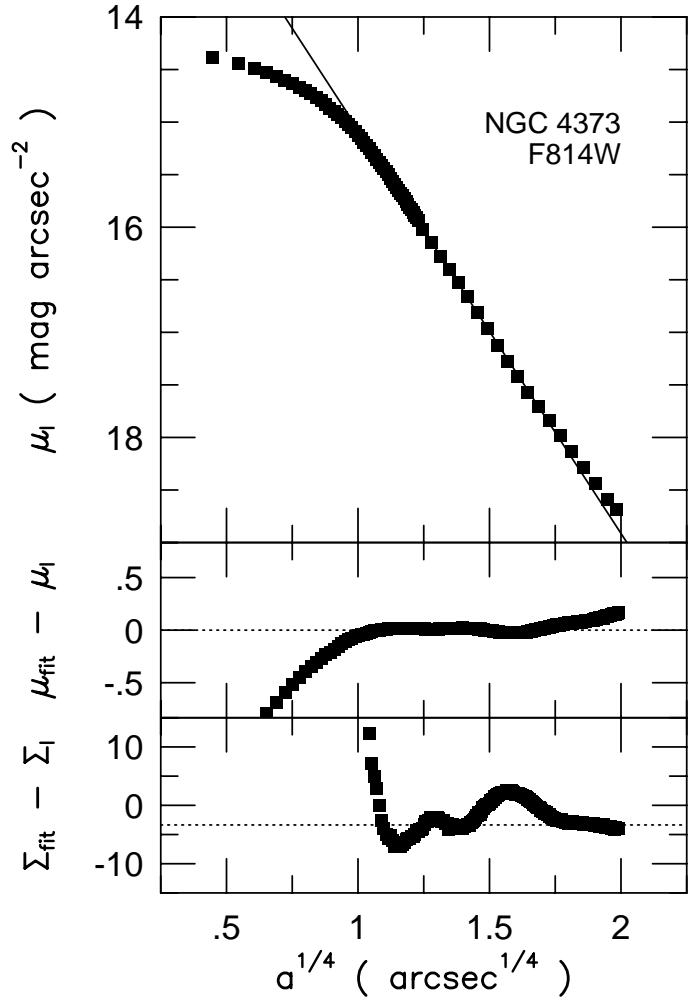}
\caption{
Surface brightness profile with fit to a de Vaucouleurs $a^{1/4}$ form
for $1 < a < 6''$.  Residuals in the surface brightness from this fit are 
given in mag~arcsec$^{-2}$ (middle) and DN~pixel$^{-1}$/1000~s (bottom).
The median sky value of $3.36$~DN/pixel/1000~s (equivalent to 153~e$^{-}$ in
6500~seconds) is shown with a dotted line in the bottom panel.
\label{n4373-sbprofile}
}
\end{figure}

A first estimate of the mean galaxy profile was constructed using the ELLIPSE task
in the ISOPHOTE package of STSDAS.
For NGC~4373 only, this profile was rebinned at large semimajor axes ($a > 30$~pix) 
in order to improve the S/N, following the procedure of Djorgovski (1985\markcite{sgd85}).  
The profile was fit to an $a^{1/4}$ de Vaucouleurs form at small semimajor axes values 
($1 < a < 6$~arcsec) where the estimated ratio of sky to galaxy is $< 4$\% ; 
this fit was extrapolated to large semimajor axes values to determine the deviation
of the profile due to the sky brightness.  
The surface brightness profile, and the residuals from the $a^{1/4}$ model fit, 
are shown in Figure~\ref{n4373-sbprofile}.
\placefigure{n4373-sbprofile}

The median predicted offset due to the sky brightness was then estimated for the 
extrapolated region to be $153 \pm 10$~e$^{-}$ for the 6500~s exposure, where the 
uncertainty is the quartile-estimated scatter of the residuals under the assumption 
that they are normally-distributed.
Since it is apparent from Figure~\ref{n4373-sbprofile} that the sky value is clearly a systematic error
instead of a random one, comparisons of that sky value with other estimates can offer
a more reasonable estimate of the sky subtraction uncertainty.
We examined the WF3 stacked image and estimated a sky value of 98~e$^{-}$ in the PC1 
image.
An estimate of the sky brightness from PC1 images taken with F814W on blank 
portions of sky (i.e. several high-redshift clusters and deep field galaxy surveys) 
puts a typical sky brightness at 135--180~e$^{-}$ in 6500~s.
Finally, we used a separate data reduction package (see \S~\ref{sbf-analysis-consistency} 
below) to measure the sky value using an extrapolated de Vaucouleurs profile, resulting in a sky
value of 198~e$^{-}$.
The first estimate of 153~e$^{-}$ was therefore judged to be reasonable, and this was
subtracted from the image itself.
A new estimate of the mean galaxy profile was then constructed as before.  
The other two galaxies, NGC~3379 and NGC~4406, had a ratio of sky to galaxy 
of 3--5\% at $a= 300 {\rm~pixel~}$ (13.7~arcsec).
Accurate sky subtraction is relatively unimportant for these galaxies,
as the uncertainty of the sky subtraction will enter the fluctuations image as the 
square--root of the fractional sky contribution to the signal.

A model image was then constructed from this profile using the first through
fourth Fourier coefficients of the fit.  
The model was subtracted from the image to construct the residual image, with 
semimajor axes between $150 < a < 300$ pixels used; the inner limit was chosen to 
remove nuclear areas that are poorly fit by the elliptical model, while the outer 
limit was required by the limit of the profile-fitting procedure (the nucleus for
NGC~4373 is off--center on the PC1 FOV).
There is some large-scale structure remaining in the image, as is typical,
since elliptical models are not perfect representations of elliptical galaxy
light distributions.

Point-sources were then masked-out in the residual image, which was smoothed on a scale 
of ten times the PSF (\cite{tonry88}); this smoothed residual image was then subtracted from the 
residual image to remove the large-scale structures.  
As a result of this step in the procedure, the low wavenumbers in the power spectrum 
are compromised, and will have to be excluded from the fit to the 
fluctuations variance.
Since the PSF size is $\sim 2$~pixel FWHM, corresponding to a Gaussian width of $\sim 0.85$~pixel,
this results in a smoothing kernel of $\sim 10$~pixel for a 512~pixel Fourier Transform (FT).
The variance of this smoothing kernel in the power spectra will fall off rapidly becoming
negligible at $> 3 \, \sigma$, or $k > 30$ in the FT, and hence wavenumbers below this value
will be excluded from the fit to the residual variance (see \S~\ref{sbf-analysis-calc-sbf}).

\subsection{Identifying Point Sources \label{sbf-analysis-point-sources} }

The residual image was then searched for sources using DAOPHOT (\cite{daophot}).
The region within $a < 150$~pixels presents difficulty for the 
detection scheme in DAOPHOT, as there is still some residual large-scale
structure remaining after the galaxy subtraction.  
Hence only the region at $150 < a < 300$~pix was searched for sources, and will be
used in the fluctuations analysis.  
The threshold was set for $5 \, \sigma$, which corresponds roughly to 
$I_{cut}=25.0$~mag for NGC~4373 and $I=24.2$~mag in the other two galaxies.
Experiments were performed with different values of $I_{cut}$, which demonstrate that
the final distance measurement for NGC~4373 is altered by $< 0.04$~mag in all cases, and
$< 0.02$~mag in most cases.
A total of 52, 14, and 8 sources were identified within this annulus for 
NGC~4373, NGC~4406, and NGC~3379, respectively.
The PSF was constructed using the brightest sources detected in this annulus, and was 
determined to be fit best by an analytical expression of the ``lorentz'' form.  
The PSF fit for NGC~4373 is dominated by the brightest source near the top of the 
frame, which is presumably a foreground galactic star.

The completeness of the detection algorithm was estimated by adding sources back into 
the residual image using ADDSTAR, and then determining what fraction at a given 
magnitude are recovered.  
This procedure was done in detail for NGC~4373, and its results were rescaled for
use on the other two galaxies.
As will be seen below, the residual variance due to unidentified sources in these 
residual images is small compared with the variance of the fluctuations in all three
galaxies.
These simulations suggested that $97$\% of sources are recovered in the NGC~4373 image 
at $I=23.75$~mag, $86$\% at $I=24.25$~mag, and $38$\% at $I=24.75$~mag.  
The recovery rate is worse than is expected from the random noise due to the sky
brightness, due to the added noise of the galaxy signal and weak, large--scale 
structures in the residual image.

A globular cluster luminosity function (GCLF) was fit to the detected sources 
simultaneously with the expected number counts of background galaxies.
The latter was estimated from WFPC--2 number counts
(\cite{cowie95}):
\begin{equation}
{dN \over dm_{\rm I}} = 240 10^{0.37 (I-17)} {\rm~deg}^{-2} {\rm~mag}^{-1},
\end{equation}
which fits both the ground-based and WFPC2 number counts up to $I<26$~mag.
The GCLF was assumed to have a gaussian shape (i.e. 
$a_0 \exp \left( - \left( {I-a_1 \over \sqrt{2} a_2} \right)^2 \right)$; 
\cite{harris88}) with the center magnitude fixed.  
The assumed parameters were $M_V = -7.1$~mag, $a_2 = \sigma=1.4$~mag, and $(V-I)=1.1$ 
for the GCLF.

\begin{figure}
\epsscale{0.6}
\plotone{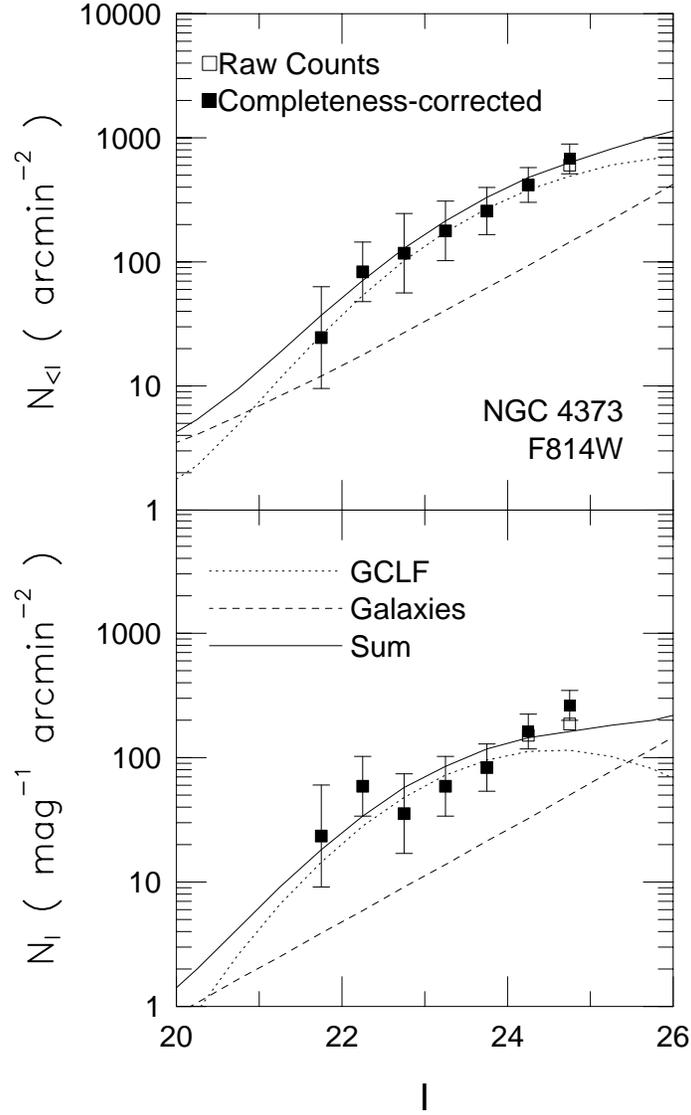}
\caption{
Luminosity function for the DAOPHOT detected objects within the annulus having
$150 < a < 300$~pixel.  Detections are shown by open symbols, while
filled symbols show the detections after correction for completeness.
The lines show the fitted model for the GCLF, galaxy number counts, and their
sum.
\label{n4373-gclf}
}
\end{figure}

A relative distance of 2.3 greater than the Virgo cluster was assumed for NGC~4373, 
while distance moduli of $31.15$ and $30.30$ were assumed for the Virgo cluster and
Leo group, respectively.
This value for the center magnitude of the GCLF is usually determined by iteration 
once a first measurement of the distance to the galaxy is complete.  
As the residual variance due to unidentified sources is quite small (see below), there
is little need to iterate to improve these first guesses for the distances.
The fit for the GCLF was done using non-linear iteration
of the Levenberg-Marquardt Method, as implemented by Press \etal\ (1986\markcite{recipes}).
Convergence was reached rapidly, producing the fitted value for the
amplitude of the gaussian.
The cutoff for the detected sources was taken as $I=25.0$~mag for NGC~4373, and $I=24.0$~mag
for the other two galaxies.
Uncertainties were included in the fit according to Poisson number statistics, with detected 
sources ranging between 1 and 12 per $0.5$~mag bin.  
The detected point sources from DAOPHOT and the fitted GCLF are shown in Figure~\ref{n4373-gclf} 
for NGC~4373.
\placefigure{n4373-gclf}

\subsection{Calculating the Fluctuation Signal \label{sbf-analysis-calc-sbf} }

The next step in the SBF analysis is to create a mask for the image.
All detected sources are masked up to a chosen limiting magnitude $I_{cut}=25.0$~mag.
Various annuli are also chosen, which here we shall take only
$150 < a < 200$~pixel, $200 < a < 250$~pixel, and $250 < a < 300$~pixel,
where the semimajor axis $a$ is chosen rather than radius in order to
construct an annulus of uniform noise.
The residual image, with the large-scale structures removed as described
above, is then divided by the square-root of the mean galaxy image
to normalize the fluctuations in each pixel.  
This is now the masked fluctuations image.  

A bright version of the PSF was constructed from DAOPHOT as described above; 
it was placed into a copy of each of the masked fluctuations images, in order that the
window function used for the fluctuations and the PSF are identical.
We compared the results of the above method to an alternate approach (\cite{tonry88}) of 
calculating the FT of the window function and introducing this FT separately into the
power spectrum of the fluctuations.
The differences between these two approaches were found to be negligible.

We note that it is certainly possible to construct a PSF for WFPC--2 images using the
``Tiny Tim'' package, but we strongly prefer the use of an empirical PSF from the
same image as the galaxy as any subtle telescope jitters or defocussing will be accounted
for properly.
A region of $512 \times 512$~pixels was then extracted from each fluctuations image and 
PSF image, and the power spectrum is calculated.
Circularly--symmetric, azimuthally--averaged radial profiles about wavenumber $k=0$ were 
then extracted from the power spectra of the fluctuations and the PSF.

The azimuthally-averaged power spectra for the luminosity fluctuations
are weighted by the inverse of the fraction of each image that is nonzero
in the mask.
The azimuthally-averaged power spectra for the PSF are normalized 
so that each represents exactly $1$~e$^{-}$ within an aperture of 
$r=11$~pixel~$= 0.50$~arcsec.
This is a crucial step, as the calibration of Holtzman \etal\ (1995b\markcite{holtzman95b}) is normalized
for the PSF in a circular aperture of precisely this radius.
Thus we can use the zero--point from Holtzman \etal\ directly via Eq.~1.
Each fluctuations power spectrum $E(k)$ is represented by:
\begin{equation}
E(k) = P_0 \times E_{psf}(k) + P_1,
\end{equation}
where $P_0$ represents the multiplicative factor for the PSF that is
the distance-dependent fluctuations magnitude, and $P_1$ is the
flat power spectrum of the Gaussian ``white'' noise.
A linear least-squares fit is performed to the data for wavenumbers
$30 < k < 256$, since a step in the reduction process (noted above) compromises
the wavenumbers $k<30$.  

\subsection{Removal of Variance Due to Unidentified Sources \label{sbf-analysis-residual-variance} }

The variance expected from point-sources not detected in the residual
image must be subtracted individually from each measurement of $P_0$.  
By integrating the variance due to undetected sources both brighter than $I_{cut}$ 
(due to incompleteness) and fainter than $I_{cut}$ using the GCLF measured in \S~\ref{sbf-analysis-point-sources}, 
the result is the residual variance $P_{res}$ for each annulus.\footnote{While the 
detection algorithm of DAOPHOT may not be 100\% complete in detecting $5 \, \sigma$
\emph{objects} in the residual image (\cite{sodemann95}),
it is 100\% complete at detecting $5 \, sigma$ \emph{peaks} in the image and hence
the latter are used for the correction of the residual variance.
Since the WFPC-2 data are very deep in their removal of point sources, reaching $I_{cut} = 25.0$~mag
at the $5 \, \sigma$ CL, any incompleteness in detecting these peaks is irrelevant to the calculation
of the residual variance.
Ground-based imaging, on the other hand, will generally not reach these faint flux limits in the
detection of point sources, so the effects of incompleteness in detecting these $5 \, \sigma$ \emph{peaks}
on the residual \emph{variance} become relevant at significant cosmological distances (and faint flux limits)
and must be estimated carefully.}
Because the correction for the residual variance due to unidentified sources is small
compared to the fluctuation signal, we did not attempt to fit the data in any other
method that would not require binning the detected sources.
Residual variance due to undetected sources was estimated by following the GCLF and 
extragalactic number counts to large magnitudes, and the result was found to converge 
very quickly at $I>25$~mag.  

Since there is a limited number of sources detected throughout the entire annulus of 
$150<a<300$~pix, it would be very challenging to estimate the GCLF for smaller 
annuli (i.e. separately for $150<a<200$, $200<a<250$, etc.).  
The GCLF is expected to vary with projected radius from the galaxy nucleus, so it is 
clearly not constant across the entire area.
Inspection of the radial dependence of the GCLF in well--studied galaxies
(such as \cite{cohen88} for NGC~4406) suggests that this effect is virtually negligible
for the small range in radii we have studied here.
The corrected SBF power is then simply $P_0^{corr} = P_0 - P_{res}$.

\subsection{Consistency Checks \label{sbf-analysis-consistency} }

A number of consistency checks were investigated for the NGC~4373 data.
First, the first wavenumber in the fit was varied between $k=30$ and $k=70$, since the 
lowest wavenumbers contribute the largest effect to the $\chi^2$ minimization routine.
Second, $P_1$ was fit for the largest wavenumbers ($225 < k < 256$) and fixed to
this value; the implied measurement of $P_0$ was then calculated for each individual 
data--point in the range $30 < k < 70$.
Both of these tests suggest that there are no significant individual outliers in the
power spectra data, nor are there systematic errors arising from a particular range of
wavenumbers.
Furthermore, these tests demonstrate that we have identified the appropriate 
minimum wavenumber for the fits, as lower wavenumbers (at $k<30$) show clear systematic
differences from the remainder of the power spectra.

The effects of the estimate of the sky value were investigated for NGC~4373, the galaxy
for which the uncertainty will be greatest, by using the different estimates described above 
in \S~\ref{sbf-analysis-model} above.
The extreme estimates of the sky value result in a change of $\pm 0.02$~mag in the
distance estimate, such that the lower sky value produces the further distance estimate.

The entire fluctuations analysis was repeated using an independent software package, 
written by J. Tonry and provided to A. Dressler.
This package can be considered the ``standard'' for the field (see \cite{tonry97},
and references therein), and was used for the ground--based $I_{\rm C}$--band distance
measurements of galaxies in the Hydra--Centaurus supercluster by Dressler (1993\markcite{dressler93}).
There are a number of significant differences between the Tonry package and the
one we have utilized above.  
One, the Tonry package finds sources using a modified DOPHOT package (\cite{dophot89}; 
\cite{dophot93}) that includes the surface brightness fluctuations in the noise 
model for detecting sources.  
Two, the completeness at 5~$\sigma$ is taken in the Tonry package to be 100\%, 
which was verified in simulations (J. Tonry, private communication); we have 
explicitly measured the completeness via simulations and account for any incompleteness 
near $I_{cut}$ in the variance due to unidentified sources (which follows closely the
method of \cite{sodemann95}).
Both of these points, however, are relatively unimportant, as the correction itself
for the residual variance is small even for NGC~4373.
Three, the Tonry package uses an analytical model fit for the power spectrum of the 
PSF instead of the power spectrum itself (i.e. a fully empirical PSF) that we have 
used above.
Four, the image regions used for the Tonry package are somewhat different from those
used above.
Comparisons were therefore made for the identical subregions of the image, namely circular 
annuli of 32:64 pixels, 64:128 pixels, and 128:256 pixels.
Fifth, the Tonry package in this version normalizes the PSF flux to effectively infinite
aperture, while our analysis above normalized the PSF flux to unity within a $r=0.5$~arcsec
aperture.
Using the encircled energy curves for the F814W filter from Holtzman \etal\ (1995a\markcite{holtzman95a}), the
difference is $\sim 9$\% in intensity, which we have applied to the Tonry package for the purposes
of this comparison.\footnote{Ajhar \etal\ (1997\markcite{ajhar97}) added a $0.1$~mag term for this same reason in calibrating 
their SBF analysis using the Tonry package (i.e. their Equation~2).}
Given the subtle but significant differences between these two analyses of the same image, the 
differences in $P_0$ are only $\sim 2$\% for NGC~4373.
This is an acceptable systematic error which is substantially smaller than the overall random
uncertainties for the SBF method (\cite{tonry97}).

\subsection{Converting $\I814bar$ Into Distances \label{sbf-analysis-calibration} }

Ajhar \etal\ (1997\markcite{ajhar97}) report a calibration of $\MI814bar$ based on detailed comparisons with
the large, ground--based study of $I_{\rm C}$--band SBF of Tonry \etal\ (1997\markcite{tonry97}).
As was shown by Tonry (1991\markcite{tonry91}), there exists a dependence of $\MIbar$ on
intrinsic color $(V-I)_0$, which was modelled using simple stellar populations 
by Worthey (1994\markcite{worthey94}).
The subsequent primary calibration of the SBF distance scale was determined by
Ajhar \etal\ (1997\markcite{ajhar97}) to be
\begin{equation}
\MI814bar = -1.73 (\pm 0.07) + 6.5 (\pm 0.7) \left[ \left( V - I \right)_0 - 1.15 \right] .
\end{equation}
The reddening estimates listed in Table~\ref{table1} are used to convert $(V-I)_{obs}$ into
$(V-I)_0$ for the three galaxies.

There is an additional cosmological correction ought to be applied.
The $k$--correction for fluctuations observed in the ground--based
$I_{\rm C}$--band filter is $7.0 \times z$ (\cite{tonry97}), which act to
brighten the observed magnitudes.
We note that SBF magnitudes act as luminosities, not surface brightnesses, hence
there is no need for additional SB dimming cosmological corrections.
Furthermore, the broadband color $(V-I)_{\rm obs}$ needs to be made bluer by
$0.9 z$.
Since these two effects nearly cancel each other in Equation~5, the possible systematic 
error on the distance modulus for NGC~4373 implied by not making these corrections is 
only $0.014$~mag, which is comparable to the uncertainties of the $k$--corrections themselves
(and their applicability to the F814W filter).
For this reason, $k$--corrections were not applied in the photometric calibration.

\section{Results \label{results} }

\subsection{NGC~3379 in the Leo Group \label{results-n3379} }

\begin{figure}
\epsscale{0.6}
\plotone{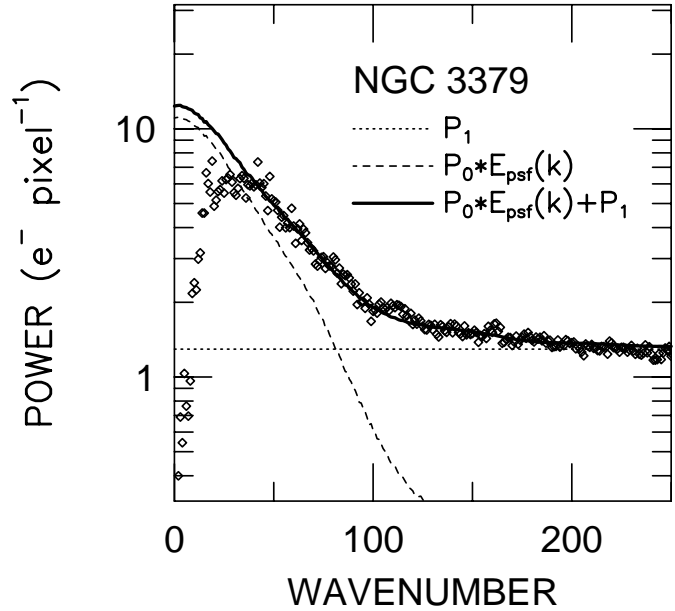}
\caption{
The power spectrum of the fluctuations signal in the $200 < a < 250$~pixel
annulus for NGC~3379.  
The solid line is the fit to the data points using the sum of two terms:  
a constant $P_1$, and a constant $P_0$ multiplied by the power spectrum 
$E_{\rm epsf}(k)$ of the PSF.
The data points at low wavenumber $k < 30$ were compromised due to a step in the
data reduction method (see \S~2), and hence only the data for $30 < k < 256$ were used
for the fit.
\label{n3379-powerspectrum}
}
\end{figure}

The fit for the intermediate annulus ($200 < a < 250$~pixel) in NGC~3379 is displayed
in Figure~\ref{n3379-powerspectrum}, and the fits for $P_0$, $P_1$, and $P_{res}$ are listed 
in Table~\ref{table2}.
\placefigure{n3379-powerspectrum}
\placetable{table2}
For the three annuli, the mean value after correction for the residual variance 
due to unmasked sources is $P_0^{corr} = 10.8 \pm 0.5$~e$^{-}$ 
in the 1340~s exposure, where the uncertainty reflects the scatter among the three
measurements divided by $\sqrt{3}$.
An estimate of the signal--to--noise of a fluctuations measurement is provided by the 
ratio $\eta = P_0/P_1$, which we measure to range between $\eta = 3.7$ and $8.6$ for 
these annuli.

While the results listed in Table~\ref{table2} suggest that there might be evidence for radial
gradients for our measurements, we caution against this interpretation without a
more thorough investigation of the HST images of this galaxy for all possible radii
obtainable with the PC array.
We have merely chosen to analyze NGC~3379 for only those three annuli identical to 
those used for NGC~4373 below, as a comparison sample.
The scatter among the three individual measurements listed in Table~\ref{table2} is $0.11$~mag,
which is comparable to the scatter in $\Ibar$ among the entire SBF survey of 
Tonry \etal\ (1997\markcite{tonry97}) of $\sim 0.1$~mag.
Thus, if a gradient exists it would only be significant at the $2$~$\sigma$~CL.
An opposite gradient in $\Ibar$ was detected by Sodemann \& Thomsen (1995\markcite{sodemann95}) for NGC~3379,
although for annuli at much larger radii ($13.5$ to $48$~arcsec) than we have studied here.
An extrapolation of their results implies that we should detect a \emph{brightening} in 
$\Ibar$ of only $0.06$~mag between the innermost and outermost annuli in the PC image.
A color gradient in $(B-J)$ but not $(J-K)$ was detected by Peletier, Valentijn, \& Jameson 
(1990\markcite{peletier90b}), so it is unclear if a gradient in $\Ibar$ should be expected
on the basis of color gradients in NGC~3379.
Errors in sky subtraction are not nearly large enough to cause the observed effect (see \S~\ref{sbf-analysis-model}).
Finally, we note that the outermost annulus had a disproportionate number of identified
points sources compared to the other annuli, which could cause a subtle measurement bias.
More detailed study is required if a fluctuations gradient is to be inferred from the
data.

The mean measurement for the fluctuations magnitude for NGC~3379 is 
$\I814bar = 28.81 \pm 0.11$~mag (after correcting for Galactic extinction) which results in a 
distance modulus of $(m - M)_0 = 30.14 \pm 0.15$~mag through application of Equation~5.
The uncertainty estimate in $\I814bar$ only represents three uncertainties:  the scatter between 
the measurements for the different regions of the image ($0.11$~mag), the stability of
the WFPC--2 zero--point for F814W of $0.018$~mag (\cite{ajhar97}),
and the 14\% uncertainty in the conversion between IRAS 100$\mu$m flux and $A_B$ extinction.
The uncertainty estimate for $\DM$ includes the uncertainty of $0.01$~mag for 
$(V-I_{\rm C})_{\rm obs}$, the uncertainty of $0.055$~mag for the fit between $(V-I_{\rm C})_0$ 
and $\MI814bar$, and $0.05$~mag to allow for cosmic scatter (\cite{tonry97}).
Our measurement of $\DM$ agrees with the value of $30.08 \pm 0.08$ from the Tonry \etal\
(1997\markcite{tonry97}) ground--based SBF survey, as quoted by Ajhar \etal\ (1997\markcite{ajhar97}).
It is somewhat nearer, however, than the value derived independently from the WFPC--2 data
by Ajhar \etal\ (significant at the $< 2 \, \sigma$ CL after correction for the different 
assumptions of Galactic extinction).

\subsection{NGC~4406 in the Virgo Cluster \label{results-n4406} }

\begin{figure}
\epsscale{0.6}
\plotone{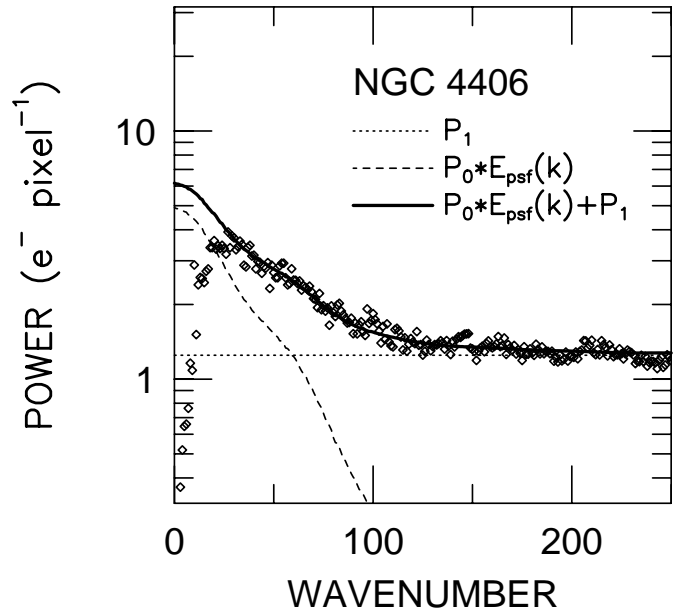}
\caption{
The power spectrum of the fluctuations signal in the $200 < a < 250$~pixel
annulus for NGC~4406.  
This figure is otherwise identical to Figure~4.
\label{n4406-powerspectrum}
}
\end{figure}

The fit for the intermediate annulus ($200 < a < 250$~pixel) in NGC~4406 is displayed
in Figure~\ref{n4406-powerspectrum}, and the fits for $P_0$, $P_1$, and $P_{res}$ are listed in Table~\ref{table2}.
\placefigure{n4406-powerspectrum}
The scatter among the different annuli is remarkably small, with no evidence for 
a radial dependence for $P_0$.
As this exposure time was similar to that of NGC~3379, but this galaxy is in the more
distant Virgo cluster, the S/N is poorer, ranging between $\eta = 1.1$ and $3.9$.
Nonetheless, the small scatter among the annuli suggests that there are no systematic
biases entering into these data as a function of $\eta$.

The mean measurement for the fluctuations magnitude for NGC~4406 is
$\I814bar = 29.83 \pm 0.03$~mag (corrected for Galactic extinction) which results in a 
distance modulus of $\DM = 31.20 \pm 0.10$.
The uncertainties were estimated in the same manner as for NGC~3379 in \S~\ref{results-n3379} above.
This distance modulus agrees with the value $31.19 \pm 0.07$~mag with the Tonry \etal\ (1997\markcite{tonry97}) 
ground--based SBF measurement as quoted by Ajhar \etal\ (1997\markcite{ajhar97}).

\subsection{NGC~4373 in the Hydra--Centaurus Supercluster \label{results-n4373} }

\begin{figure}
\epsscale{0.4}
\plotone{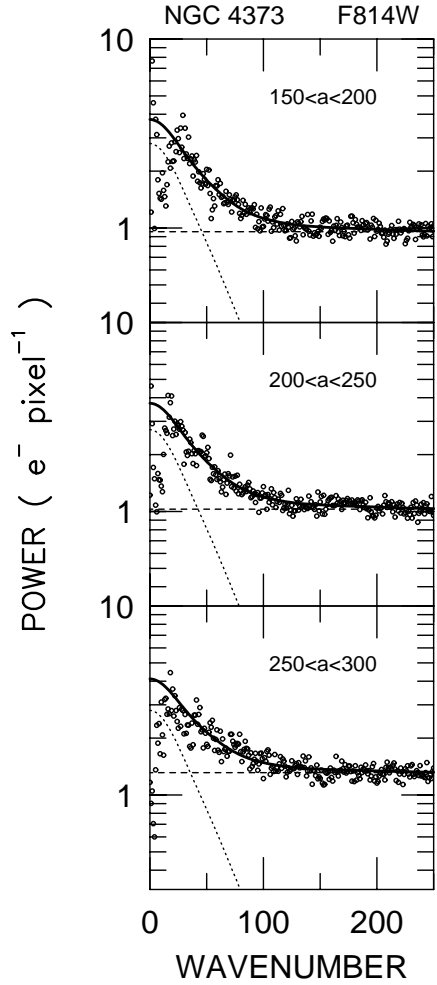}
\caption{
Power spectra of the surface brightness fluctuations within
three different annuli from the F814W image of NGC~4373.
The inner and outer semimajor axes of each annulus is indicated
in the upper-right portion of each panel.  The data are given
by open circles, which are fit as the sum (heavy solid line) 
of $P_0$ times the PSF (dotted lines) and $P_1$ (dashed line).
The gaussian noise ($P_1$) can be seen to increase with semimajor
axis, as expected because the fluctuations have been normalized by
dividing by the square-root of the mean galaxy flux.
\label{n4373-powerspectra}
}
\end{figure}

The fits to each of the three annuli are displayed in Figure~\ref{n4373-powerspectra}, and the results are
listed in Table~\ref{table2}.
\placefigure{n4373-powerspectra}
The results from the three annuli that are mutually consistent at the $0.02$~mag level.
This exposure was longer (6500~s) than either of the other two galaxies, but is
further in distance, hence the S/N is comparable to the best of NGC~4406, i.e. 
$\eta = 3.6$ to $4.8$.

The mean measurement for the fluctuations magnitude for NGC~4373 is
$\I814bar = 31.31 \pm 0.05$~mag (corrected for Galactic extinction) which results in a 
distance modulus of $\DM = 32.99 \pm 0.11$.
Once again, the uncertainties are estimated as described in \S~\ref{results-n3379}, with an additional
$0.02$~mag added due to sky--subtraction uncertainties that are not relevant for
the other two galaxies.

Dressler \etal\ (1997, in preparation) have made new measurements of the distances
to Hydra--Centaurus galaxies using ground--based SBF data obtained under excellent seeing
conditions.
Preliminary results described by Dressler (1993\markcite{dressler93}) for NGC~4373 were obtained with relatively
poor seeing ($1.03$~arcsec FWHM) compared to the newer data ($0.76$~arcsec FWHM), hence these
new measurements supersede the older ones.
Using the calibration of Tonry \etal\ (1997\markcite{tonry97}) for these new ground--based data and the same extinction
as listed in Table~\ref{table1}, the preliminary distance modulus is $32.93$, in excellent agreement with 
the value derived from the WFPC--2 observations.

Peculiar velocities should be calculated relative to the baseline expansion rate of
$H_0 = 81 \pm 6$~km~s$^{-1}$~Mpc$^{-1}$ derived by Tonry \etal\ (1997\markcite{tonry97}) from their SBF database,
since their study was used in the present paper for the calibration of 
$\MI814bar$.\footnote{The Ajhar \etal\ (1997\markcite{ajhar97}) SBF calibration for HST filter F814W was done 
directly onto the Tonry \etal\ (1997\markcite{tonry97}) system by assuming the distance moduli for each galaxy
from the latter study.  Hence the two are equivalent for our purpose.}
Taking the velocity of the NGC~4373 group with respect to the CMB as $3625$~km~s$^{-1}$ 
(\cite{faber89}), then the distance modulus of $32.99 \pm 0.11$ suggests that NGC~4373 possesses
a peculiar velocity of $v_{\rm pec} = 415 \pm 330$~km~s$^{-1}$ with respect to the CMB.
Alternatively, the ratio of distances between NGC~4373 and NGC~4406 can be compared to the
ratio of the CMB relative velocities of the NGC~4373 group and the Virgo cluster (included in Table~\ref{table1})
to calculate a peculiar velocity for NGC~4373.
This separate calculation yields a peculiar velocity of $v_{\rm pec} = 220 \pm 270$~km~s$^{-1}$.
The cause of the difference between these two estimates of $v_{\rm pec}$ is the somewhat high value 
of the cosmological redshift in the CMB frame for the Virgo cluster ($1493$~km~s$^{-1}$, from \cite{faber89}); 
taking the SBF value of this quantity ($1390$~km~s$^{-1}$, from \cite{tonry95}) results in 
$v_{\rm pec} = 455 \pm 270$~km~s$^{-1}$.
We adopt the value of $v_{\rm pec} = 415 \pm 330$~km~s$^{-1}$, as it does not require the assumption
of the CMB velocity of the Virgo cluster from the $D_n$--$\sigma$ relation.

\section{Discussion \label{discussion} }

\subsection{Feasibility of SBF With HST \label{discussion-feasibility} }

This paper represents the detection of surface brightness fluctuations
in images taken with the WFPC--2 instrument on HST.
The images are marginally sampled, or slightly under--sampled, even using the
Planetary Camera.
Nonetheless, we have demonstrated that the fluctuations signal is strong with only
a modest exposure time at a distance of the NGC~4373 group in the Hydra--Centaurus
supercluster.

The measurements of $\DM$ in NGC~3379 and NGC~4406 agree well with 
ground--based measurements, suggesting that there are no substantial systematic 
offsets between the marginally--sampled HST SBF data and the oversampled
ground--based work at uncertainties of less than 10\% in distance.
The Planetary Camera of HST is thus a potentially powerful tool for measuring
distances using the SBF method.

The removal of point sources (foreground stars and globular clusters) and background 
galaxies is straightforward.
The estimation of the residual variance due to other undetected globular clusters and
galaxies is robust, and the correction implied is quite small, typically 2--4\%.
This is due to the combination of the depth afforded by the high--resolution imaging
power of HST, and by the small projected areas that have been studied here.
Ground--based imaging typically utilizes a much larger FOV for the calculation of
$\Ibar$, resulting in a larger number of undetected objects that must be
removed from the SBF signal.

Estimating foreground Galactic extinction is an important element of SBF work with HST,
as the slope of the dependence of $\MI814bar$ on $(V-I_{\rm C})_0$ is 6.5.
This causes errors in the distance modulus to be $\Delta \DM \sim 1.5 \Delta A_B$, 
which might seem surprising since the SBF observations are made at $\lambda = 800$~nm.
The minimal effect of $k$--corrections on $(\overline{m} - \overline{M})$, however, 
does provide a reduction in additional sources of uncertainty.

\subsection{The Distance to NGC~4373 \label{discussion-n4373-distance} }

Our distance measurement to NGC~4373 ought to be more accurate than ground--based work
due to the greater confidence at which we can estimate $P_{res}$.
Experiments with varying values of $I_{cut}$ and the S/N for detecting such sources
suggests that the possible variations in $P_{res}$ are $\leq 4$\% of $P_0$.
Comparison of two different SBF analysis packages suggests that the systematic errors
due to methodology are $\simlt 2$\% in distance, possibly much less if we account 
for differences in the normalization of the PSF and sky subtraction.

\begin{figure}
\epsscale{0.6}
\plotone{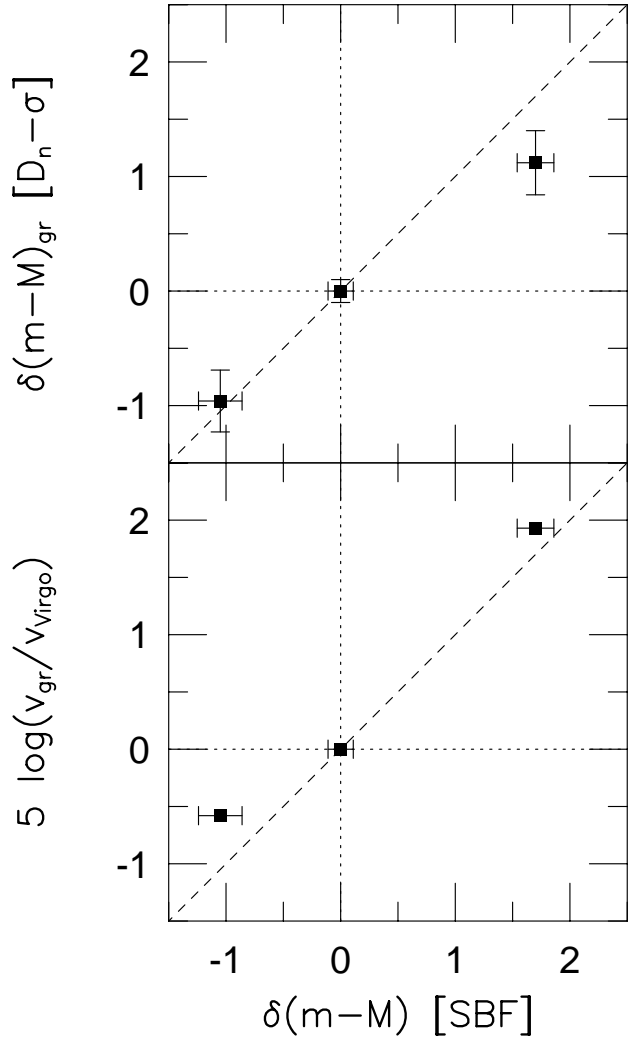}
\caption{
[top] Comparison of the SBF distances estimates for the three galaxies with those
of each group or cluster using the $D_n$--$\sigma$ relation (\cite{faber89}).
[bottom] Comparison of the SBF distances estimates for the three galaxies with those
of each group or cluster expected from the group velocity (from \cite{faber89}).
In both panels, the distances and velocities have been normalized to the Virgo cluster.
As seen in the top panel, the SBF distance to NGC~4373 is slightly larger than that
from the $D_n$--$\sigma$ relation; part of this difference could be due to NGC~4373
being further than the other members of its group, as suggested by the ground--based
SBF measurements of Dressler (1993).
\label{compare-distances}
}
\end{figure}

The distance we estimate for NGC~4373 implies a smaller peculiar velocity than has
been measured previously.
A comparison of the SBF distance estimates with those of the $D_n$--$\sigma$ relation
(\cite{faber89}) and the group velocities (with respect to the CMB) are shown in
Figure~\ref{compare-distances}, where the distances and velocities have been calculated 
relative to the Virgo cluster.
\placefigure{compare-distances}
The source of the large peculiar velocity of the Hydra--Centaurus supercluster
is shown by POTENT m.eps as a mass concentration at a redshift of 
$\sim$4500 km/sec (\cite{dekel95}). 
The model for the velocity flows in this direction based on $D_n$--$\sigma$ observations 
(\cite{lb88}) predicts a peculiar velocity of 838~km~s$^{-1}$ for NGC~4373.
Based on Tully--Fisher distance measurements of clusters in the Hydra--Centaurus region, however,
Aaronson \etal\ (1989\markcite{aaronson89}) and Mould \etal\ (1991\markcite{mould91}) measured the median value of the peculiar 
velocity flow to be 489 and 610~km~s$^{-1}$, respectively.
Our measured peculiar velocity of $415 \pm 330$~km~s$^{-1}$ for NGC~4373 is therefore in better 
agreement with the values derived from the Tully--Fisher method, although the measurement is only 
in marginal conflict with the model based on the $D_n$--$\sigma$ data.
We note that the peculiar velocity to the NGC~4373 group might be somewhat
larger if IC~3370 and NGC~4373A are closer than NGC~4373, as implied by the 
ground--based SBF results of Dressler (1993\markcite{dressler93}), although this result ought to be reinvestigated 
using observations obtained in better seeing conditions.

It is clear from this simple analysis that more SBF measurements of distances, both ground--based 
and HST, should be attempted to understand better the velocity flows in the Hydra--Centaurus
region.

\acknowledgments

We acknowledge many helpful discussions with J. Blakeslee and G. Worthey, and the insightful
comments and careful reading of the manuscript by the referee, J. Tonry.
This work was supported by NASA grant NAS7--1260 to the WFPC--2 IDT.
M.~A.~P. received partial support from NSF grant AST--9157412 and Hubble Fellowship grant 
HF-01099.01-97A from STScI (which is operated by AURA under NASA contract NAS5-26555).

\clearpage
 

\makeatletter
\def\jnl@aj{AJ}
\ifx\revtex@jnl\jnl@aj\let\tablebreak=\nl\fi
\makeatother

\begin{deluxetable}{lcccccccccc}
\tablewidth{0pc}

\tablecaption{
Observed and Derived Quantities for the Galaxies
\label{table1}
}

\scriptsize

\tablehead{
\colhead{Galaxy} & \colhead{$v_{gr}^{CMB}$} & \colhead{$F_{100}$}       & \colhead{$A_B$} & \colhead{$E(V-I_{\rm C})$} & 
\colhead{$A_{F814W}$} & \colhead{$(V-I_{\rm C})_{obs}$}	& \colhead{$(V-I_{\rm C})_0$} & \colhead{t} & \colhead{$\Io814bar$} &
\colhead{$(m-M)_0$}
\\
\colhead{~}	 & \colhead{(km~s$^{-1}$)}  & \colhead{(MJy~sr$^{-1}$)} & \colhead{(mag)} & \colhead{(mag)} &
\colhead{(mag)}       & \colhead{(mag)}                 & \colhead{(mag)}             & \colhead{(s)} & \colhead{(mag)} &
\colhead{(mag)}
\\
\colhead{(1)}	& \colhead{(2)}	& \colhead{(3)}	& \colhead{(4)} & \colhead{(5)} & \colhead{(6)}	& \colhead{(7)} 
& \colhead{(8)} & \colhead{(9)} & \colhead{(10)} & \colhead{(11)} 
}

\startdata

NGC~3379 & 1142 & 2.3 & 0.164 & 0.049 & 0.079 & 1.261 & 1.212 & 1340 & 28.81 & 30.14 \nl
NGC~4406 & 1493 & 2.4 & 0.171 & 0.051 & 0.082 & 1.257 & 1.206 & 1500 & 29.83 & 31.20 \nl
NGC~4373 & 3625 & 6.7 & 0.479 & 0.144 & 0.229 & 1.302 & 1.158 & 6500 & 31.31 & 32.99 \nl

\tablecomments{Column (2) is from Faber \etal\ (1989).  The $(V-I_{\rm C})$ colors
	for NGC~3379 and NGC~4406 are taken from Ajhar \etal\ (1997) and corrected to
	match the extinction in column (5), while $(V-I_{\rm C})$ for NGC~4373 is
	taken from Dressler \etal\ (in preparation).  Measurements of $\I814bar$ in column (11)
	are taken from Table~2 along with the correction for Galactic extinction from column (6). }

\enddata

\normalsize

\end{deluxetable}

\clearpage


\makeatletter
\def\jnl@aj{AJ}
\ifx\revtex@jnl\jnl@aj\let\tablebreak=\nl\fi
\makeatother

\begin{deluxetable}{lccccccc}
\tablewidth{39pc}

\tablecaption{
Fluctuations Analysis
\label{table2}
}

\tablehead{
\colhead{Galaxy}		&
\colhead{Annulus}		& \colhead{$P_0$}		& 
\colhead{$P_1$}			& \colhead{$P_{res}$}	&
\colhead{$P_0^{corr}$}		& \colhead{$\I814bar$}	& \colhead{$\sigma_{\I814bar}$}
\\
\colhead{~}			&
\colhead{(pixel)}		& 
\colhead{(e$^{-}$~pix$^{-1}$)}	& 
\colhead{(e$^{-}$~pix$^{-1}$)}	& \colhead{(e$^{-}$~pix$^{-1}$)}	&
\colhead{(e$^{-}$~pix$^{-1}$)}	& \colhead{(mag)}		& \colhead{(mag)}
}

\startdata


NGC~3379  &  150:200  &  11.81  &   1.56  &   0.04  &  11.77  &  28.80  &   0.02       \nl
\nodata   &  200:250  &  11.16  &   1.30  &   0.05  &  11.11  &  28.86  &   0.04       \nl
\nodata   &  250:300  &   9.68  &   2.64  &   0.05  &   9.63  &  29.02  &   0.04       \nl
\nodata   &   mean    & \nodata & \nodata & \nodata &  10.83  &  28.89  &   0.11       \nl
\nl
NGC~4406  &  150:200  &   4.68  &   1.22  &   0.04  &   4.64  &  29.93  &   0.05       \nl
\nodata   &  200:250  &   4.88  &   1.25  &   0.05  &   4.83  &  29.89  &   0.02       \nl
\nodata   &  250:300  &   4.78  &   4.40  &   0.06  &   4.72  &  29.91  &   0.05       \nl
\nodata   &   mean    & \nodata & \nodata & \nodata &   4.73  &  29.91  &   0.02       \nl
\nl
NGC~4373  &  150:200  &   4.61  &   0.96  &   0.05  &   4.56  &  31.54  &   0.02       \nl
\nodata   &  200:250  &   4.57  &   1.03  &   0.06  &   4.51  &  31.56  &   0.05       \nl
\nodata   &  250:300  &   4.75  &   1.31  &   0.08  &   4.67  &  31.52  &   0.04       \nl
\nodata   &   mean    & \nodata & \nodata & \nodata &   4.58  &  31.54  &   0.02       \nl

\enddata

\end{deluxetable}

\clearpage


\clearpage


\begin{thebibliography}{}

\bibitem[Aaronson \etal\ 1989]{aaronson89}
	Aaronson, M., \etal\  1989, \apj, 338, 654
\bibitem[Ajhar \etal\ 1997]{ajhar97}
	Ajhar, E. A., Lauer, T. R., Tonry, J. L., Blakeslee, J. P.,
	Dressler, A., Holtzman, J. A., \& Postman, M.
	1997, \aj, 114, 626
\bibitem[Blakeslee \& Tonry 1995]{blakeslee95}
	Blakeslee, J. P., \& Tonry, J. L.  1995, \apj, 442, 579
\bibitem[Blakeslee 1997]{blakeslee97}
	Blakeslee, J. P.  1997, \apj, 481, L59
\bibitem[Cardelli, Blayton, \& Mathis 1989]{cardelli89}
	Cardelli, J. A., Clayton, G. C., \& Mathis, J. S.  1989, \apj, 345, 245
\bibitem[Cohen 1988]{cohen88}
	Cohen, J. G.  1988, \aj, 95, 682
\bibitem[Cowie, Hu, \& Songaila 1995]{cowie95}
	Cowie, L. L., Hu, E. M., \& Songaila, A.  1995, \aj, 110, 1576
\bibitem[Dekel 1995]{dekel95}
	Dekel, A.  1995, Proceedings of the Heron Island Workshop
	on Peculiar Velocities in the Universe, \verb+http://qso.lanl.gov/~heron/+
\bibitem[Djorgovski 1985]{sgd85}
	Djorgovski, S.  1985, Ph.D. Thesis, UC Berkeley
\bibitem[Dressler 1993]{dressler93}
	Dressler, A.  1993, in Cosmic Velocity Fields, Proceedings of the 9th IAP Astrophysics
	Meeting, eds. F. R. Bochet \& M. Lachi\`eze--Rey (Gif--sur--Yvette:  Editions
	Frontieres), 9
\bibitem[Faber \etal\ 1989]{faber89}
	Faber, S. M., Wegner, G., Burstein, D., Davies, R. L., Dressler, A.,
	Lynden--Bell, D., \& Terlevich, R. J.  1989, \apjs, 69, 763
\bibitem[Harris 1988]{harris88}
	Harris, W. E. 1988, in The Extragalactic Distance Scale,
	ed. S. v. d. Bergh \& C. J. Pritchet, 231
\bibitem[Holtzman \etal\ 1995a]{holtzman95a}
	Holtzman, J., \etal\  1995a, \pasp, 107, 156
\bibitem[Holtzman \etal\ 1995b]{holtzman95b}
	Holtzman, J. A., Burrows, C. J., Casertano, S., Hester, J. J., 
	Trauger, J. T., Watson, A. M., \& Worthey, G.   
	1995b, \pasp, 107, 1065
\bibitem[Jacoby \etal\ 1992]{jacoby92}
	Jacoby, G. H., \etal\  1992, \pasp, 104, 599
\bibitem[Jensen, Luppino, \& Tonry 1996]{jensen96}
	Jensen, J. B., Luppino, G. A., \& Tonry, J. L.  1996,
	\apj, 468, 519
\bibitem[Lauer \etal\ 1997]{lauer97}
	Lauer, T. R., Tonry, J. L., Postman, M., Ajhar, E. A., \& Holtzman, J. A.
	1997, \apj, in press
\bibitem[Laureijs, Helou, \& Clark 1994]{laureijs94}
	Laureijs, R. J., Helou, G., \& Clark, F. O.  1994, in Proceedings of 
	The First Symposium on the Infrared Cirrus and Diffuse Interstellar Clouds, 
	ASP Conf. Ser. Vol. 58, eds. R. M. Cutri \& W. B. Latter (San Francisco,
	ASP), 133
\bibitem[Luppino \& Tonry 1993]{lupton93}
	Luppino, G. A., \& Tonry, J. L.  1993, \apj, 410, 81
\bibitem[Lynden-Bell \etal\ 1988]{lb88}
	Lynden--Bell, D., Faber, S. M., Burstein, D., Davies, R. L.,
	Dressler, A., Terlevich, R. J., \& Wegner, G.  1988, \apj, 326, 19
\bibitem[Mateo \& Schechter 1989]{dophot89}
	Mateo, M., \& Schechter, P. L.  1989, in First ESO/ST--ECF
	Data Analysis Workshop, eds. P. J. Grosb\o l, F. Murtagh, \&
	R. H. Warmels (Garching:  ESO), 69
\bibitem[Mould \etal\ 1991]{mould91}
	Mould, J. R., Han, M., Roth, J., Staveley--Smith, L., Schommer, R. A.,
	Bothun, G. D., Hall, P. J., Huchra, J. P., Walsh, W., \& Wright, A. E.
	1991, \apj, 383, 467
\bibitem[Pahre \& Mould 1994]{pahre94}
	Pahre, M. A., \& Mould, J. R.  1994, \apj, 433, 567
\bibitem[Peletier \etal\ 1990]{peletier90b}
        Peletier, R. F., Valentijn, E. A., \& Jameson, R. F.
        1990, \aap, 233, 62   
\bibitem[Press \etal\ 1986]{recipes}
	Press, W. H., Flannery, B. P., Teukolsky, S. A., \&
	Vetterlin, W. T.  1986, Numerical Recipes (Cambridge:
	Cambridge Univ. Press)
\bibitem[Schechter, Mateo, \& Saha 1993]{dophot93}
	Schechter, P. L., Mateo, M., \& Saha, A.  1993, \pasp, 105, 1342
\bibitem[Simard \& Pritchet 1994]{simard94}
	Simard, L., \& Pritchet, C. J.  1994, \aj, 107, 503
\bibitem[Sodemann \& Thomsen 1995]{sodemann95}
	Sodemann, M., \& Thomsen, B.  1995, \aj, 110, 179
\bibitem[Stetson 1987]{daophot}
	Stetson, P. B.  1987, \pasp, 99, 191
\bibitem[Tonry \& Schneider 1988]{tonry88}
	Tonry, J., \& Schneider, D. P.  1988, \aj, 96, 807
\bibitem[Tonry, Ajhar, \& Luppino 1989]{tal89}
	Tonry, J. L., Ajhar, E. A., \& Luppino, G. A.  1989,
	\apj, 346, L57
\bibitem[Tonry, Ajhar, \& Luppino 1990]{tal90}
	Tonry, J. L., Ajhar, E. A., \& Luppino, G. A.  1990,
	\aj, 100, 1416
\bibitem[Tonry \& Schechter 1990]{tonsch90}
	Tonry, J. L., \& Schechter, P. L.  1990, \aj, 100, 1794
\bibitem[Tonry 1991]{tonry91}
	Tonry, J. L.  1991, \apj, 373, L1
\bibitem[Tonry 1995]{tonry95}
	Tonry, J. L.  1995, Proceedings of the Heron Island Workshop
	on Peculiar Velocities in the Universe, \verb+http://qso.lanl.gov/~heron/+
\bibitem[Tonry \etal\ 1997]{tonry97}
	Tonry, J. L., Blakeslee, J. P., Ajhar, E. A., \&
	Dressler, A.  1997, \apj, in press
\bibitem[Worthey 1994]{worthey94}
	Worthey, W.  1994, ApJS, 95, 107

\end{thebibliography}
\end{document}